\documentstyle[a4,
twoside]{article}
\makeatletter
\renewcommand{\@oddhead}{An Interaction of An Oscillator with ...
 . II: Resolvents formulae  \hfill \thepage}
\renewcommand{\@evenhead}{\thepage \hfill S.A. Choro\v{s}avin }
\renewcommand{\@oddfoot}{}
\renewcommand{\@evenfoot}{}
\makeatother

{\vspace*{2ex}\par {#1}\quad\it }{\medskip\par }

\newenvironment{Thm}[2]%
{\par\addvspace{\bigskipamount}{\bf #1#2}\it }%
{\par\addvspace{\bigskipamount} }

\newenvironment{Observation}[1]{\begin{Thm}{Observation}{#1}}{\end{Thm} }

\newenvironment{Remark}[1]{\begin{Thm}{Remark}{#1}}{\end{Thm} }

{\par\hspace*{\fill}$\Box$\par\addvspace{\bigskipamount} }

{\par\hspace*{\fill}$\Box$ \par\addvspace{\bigskipamount} }

\newcommand{\dfrac}[2]{{\displaystyle \frac{#1}{#2}}}

\newcommand{\intt}%
{
\makebox[3ex][c]
{\makebox[0pt]{$-$}\makebox[0pt]{$\displaystyle \int$}}
}

\author{S.A.~Choro\v{s}avin}
\title{ 
  An Interaction of An Oscillator with 
  An One-Dimensional Scalar Field.
 \bigskip\\
  Simple Exactly Solvable Models based on \\ 
  Finite Rank Perturbations Methods. \bigskip\\ 
  II: Resolvents formulae }
\date{}
\begin{document}
\maketitle 
\begin{abstract}
 This paper is an electronic application to my set of lectures, 
 subject:`Formal methods in solving differential equations and 
 constructing models of physical phenomena'. Addressed, mainly: 
 postgraduates and related readers. 
 Content: a very detailed discussion of the simple model of interaction based 
 on the equation array: 
$$
 z
\left(\begin{array}{cc}
  q \\
  u
\end{array}\right)
 -
\left(\begin{array}{cc}
 -\Omega^2         
       &  \Omega^2<l|_2 \\
 4\gamma_c\delta_{\alpha,x_0}<1|_1 
       &  B -4\gamma_c\delta_{\alpha,x_0}<l|_2
\end{array}\right)
\left(\begin{array}{c}
 q \\
 u
\end{array}\right)
 = 
\left(\begin{array}{cc}
 w_1 \\
 w_2
\end{array}\right)
$$
 Besides, less detailed discussion of related models.
 Central mathematical points: 
 Finite Rank Perturbations Methods, Resolvents formulae,  
 Donoghue-like models, Friedrichs-like models.
 Central physical points: phenomenon of Resonance 
 and notion of Second Sheet. 

 Hereafter I use a P.A.M. Dirac's ``bra-ket'' syntax and suppose that 
$B$ stands for an abstract linear operator,
$l$ for a linear functional, $u, w_2, \delta_{\alpha,x_0}$ 
 for abstract elements; $q, w_1 z, \Omega, \gamma_c$ stand for numbers. 
$q, u$ are objects to be found, the others are arbitrarily given.
\end{abstract}


\newpage 
\section*%
{ Introduction. } 
\subsection*%
{ A Harmonic Oscillator Coupled to an One-Dimensional Scalar Field. }

 In a previous paper I had discussed several models of 
 an one-dimensional harmonic oscillator 
 coupled to an one-dimensional scalar field. 
 Primarily I was interested in the models 
 which one can describe by the equation array

\begin{eqnarray*}
 \ddot q
 &=&-\Omega^2(q-Q)+f_0(t) 
\\
 \ddot u 
 &=&
 Bu -4\gamma_c \Big(\delta_{\alpha,t,x_0}\Big)\cdot\Big(Q-q\Big)+f_1(t) 
\\ 
 Q  
 &=& Q(t) = <l|u(t)> 
\\
\end{eqnarray*}
 as well as the models which one can describe by the equation array
\begin{eqnarray*}
 \ddot q_0
 &=&-\Omega^2 q_0 + \gamma_1 Q_{\phi} +f_0(t) 
\\
 \ddot \phi 
 &=&
 B\phi  + 4\gamma_{2,c} \Big(\delta_{\alpha,t,x_0}\Big)\cdot q_0 +f_1(t) 
\\ 
 Q_{\phi}  
 &=& Q_{\phi}(t) = <l|\phi(t)> 
\\
\end{eqnarray*}
 Hereafter I use a P.A.M. Dirac's ``bra-ket'' syntax and suppose that 
$q$ and $Q$, $q_0$ and $Q_{\phi}$
 are usual (one-dimensional) functions of 
$t$ : 
$$
 q=q(t) \,,\quad Q=Q(t) \,,\quad 
 q_0=q_0(t) \,,\quad Q_{\phi}=Q_{\phi}(t) \,, 
$$ 
$B$ 
 is an abstract linear operator,
$l$
 is a linear functional,
$\{u(t)\}_{t}$, 
$\{\phi(t)\}_{t}$
 and 
$\{\delta_{\alpha,t,x_0}\}_{t}$ 
 are families of abstract elements; 
 of course the type of 
$\delta_{\alpha,t,x_0}$ 
 must be the same as one of 
$u(t)$ or resp. 
$\{\phi(t)\}_{t}$.

 Normally I supposed that 
$\delta_{\alpha,t,x_0}$ 
 did not depend on 
$t$, 
 i.e. was constant in 
$t$. 
 In that case I wrote  
$\delta_{\alpha,x_0}$ 
 instead of 
$\delta_{\alpha,t,x_0}$.

 In this second paper I will 
 suppose that 
$\delta_{\alpha,t,x_0}$ 
 is constant in 
$t$
 as well,  
 and rewrite the above equations as follows: 
$$
\left(\begin{array}{cc}
 \ddot q \\
 \ddot u
\end{array}\right)
 =
\left(\begin{array}{cc}
 -\Omega^2         
       &  \Omega^2<l|_2 \\
 4\gamma_c\delta_{\alpha,x_0}<1|_1 
       &  B -4\gamma_c\delta_{\alpha,x_0}<l|_2
\end{array}\right)
\left(\begin{array}{c}
 q \\
 u
\end{array}\right)
 + 
\left(\begin{array}{cc}
 f_0(t) \\
 f_1(t)
\end{array}\right)
$$
$$
\left(\begin{array}{cc}
 \ddot q_0 \\
 \ddot \phi 
\end{array}\right)
 =
\left(\begin{array}{cc}
 -\Omega^2         
       &  \gamma_1<l|_2 \\
 4\gamma_{2,c}\delta_{\alpha,x_0}<1|_1 
       &  B 
\end{array}\right)
\left(\begin{array}{c}
 q_0 \\
 \phi 
\end{array}\right)
 + 
\left(\begin{array}{cc}
 f_0(t) \\
 f_1(t)
\end{array}\right)
$$
 where the subscribts 
$1$ and $2$ in 
$<\cdots|_1$ and $<\cdots|_2$ 
 mean that arguments of 
$<\cdots|_1$ 
 are elements of the first component of the vector 
$\left(\begin{array}{c}
 q \\
 u
\end{array}\right)$
 resp.
$\left(\begin{array}{c}
 q_0 \\
 \phi
\end{array}\right)$
 and arguments of 
$<\cdots|_2$ 
 are elements of the second component of the suitable vector.
\footnote{
 often we omite these subscripts, using them primarily for emphasis. 
}
 Of course, 
$$ 
 <1|q> = <1|_1 q> = q\,, \quad <1|q_0> = <1|_1 q_0> = q_0 
$$  

 The subject will primarily be 
 {\bf resolvents formulae}
 i.e. 
 the formulae that resolve the equation array:  
$$
 z
\left(\begin{array}{cc}
  q \\
  u
\end{array}\right)
 -
\left(\begin{array}{cc}
 -\Omega^2         
       &  \Omega^2<l|_2 \\
 4\gamma_c\delta_{\alpha,x_0}<1|_1 
       &  B -4\gamma_c\delta_{\alpha,x_0}<l|_2
\end{array}\right)
\left(\begin{array}{c}
 q \\
 u
\end{array}\right)
 = 
\left(\begin{array}{cc}
 w_1 \\
 w_2
\end{array}\right)
$$
 resp.
$$
 z
\left(\begin{array}{cc}
  q_0 \\
  \phi 
\end{array}\right)
 -
\left(\begin{array}{cc}
 -\Omega^2         
       &  \gamma_1<l|_2 \\
 4\gamma_{2,c}\delta_{\alpha,x_0}<1|_1 
       &  B 
\end{array}\right)
\left(\begin{array}{c}
 q_0 \\
 \phi 
\end{array}\right)
 = 
\left(\begin{array}{cc}
 w_1 \\
 w_2
\end{array}\right)
$$

 Of course, here all quantities,
$q, u, w_1, w_2,\delta_{\alpha,x_0}, l $, 
 etc. are supposed to be constant in  
$t$. 

 In the previous paper I had discussed 
 d'Alembert-like solutions to the systems and resp.
 phenomena of {\bf Radiation Reaction}, {\bf Braking Radiation} 
 and {\bf Resonance}.

 In this paper I will discuss 
 Donoghue-Friedrichs-like solutions to the systems
 \footnote{ i.e., I will discuss resolvents formulae of the system } 
 and resp.
 the phenomenon of {\bf Resonance} and notion of the {\bf Second Sheet}.

\newpage\section
{ Finite Rank Perturbations. Abstract Formulae.}  
\subsection
{ Simple Cases. Rank-One Perturbations and Rank-Two Perturbations}
\subsubsection
{ Rank-One Perturbations. }

 The simplest case of the problem looks like this:
 Suppose, we can find 
$v_0$ 
 to an equation 
$$
 Av_0=w_0
$$
 for any given 
$w_0$, 
 where 
$A$ 
 is an invertible linear operator, 
 and suppose, we need to find any 
$v$
 to 
$$
 Av-f_a<l_a|v>=w
$$
 where 
$<l_a|$ 
 is a given linear functional, and the elements 
$f_a , w$ 
 are given, as well. 

 One is used to saying concisely: we need 
$$
 (A-f_a<l_a|)^{-1}w
$$

 Then we do as follows: 

 First, write 
$$
 Av=w+f_a<l_a|v>
$$
 and notice, 

\bigskip 

\fbox{\bf 
 in order to obtain 
$v$
 we need to obtain ONLY 
$<l_a|v>$ } 
\bigskip 
\\
 Actually, 
$$
 v=A^{-1}(w+f_a<l_a|v>)
$$
$$
 v=A^{-1}w+<l_a|v>A^{-1}f_a 
$$
 So,  
$$
\fbox{ 
$\displaystyle
 v=A^{-1}w+c_a A^{-1}f_a \,,
\qquad\mbox{ where }  c_a:=<l_a|v>.
$}
$$
 Apply now this form to itself and to the recent relationship, consequently:   
$$
 v=A^{-1}w+<l_a|v>A^{-1}f_a 
$$
 then  
$$
 <l_a|v>=<l_a|\Bigl(A^{-1}w+<l_a|v>A^{-1}f_a\Bigr)> 
$$
 i.e., 
$$
 c_a=<l_a|A^{-1}w+c_a A^{-1}f_a> 
$$
 On the other hand, to find a 
$c_a$, 
 it is sufficient to fulfil  
$$
 A^{-1}w+c_a A^{-1}f_a=A^{-1}w+<l_a|A^{-1}w+c_a A^{-1}f_a>A^{-1}f_a 
$$
 i.e., 
$$
 c_a A^{-1}f_a=<l_a|A^{-1}w+c_a A^{-1}f_a>A^{-1}f_a 
$$
 This relation is fulfilled, if 
$$
 c_a=<l_a|A^{-1}w+c_a A^{-1}f_a> 
$$
 It is the same equation to 
$c_a$ 
 as above.
 Therefore, we need to solve this only scalar equation
 and set 
$ v=A^{-1}w+c_a A^{-1}f_a$ .
 To do it, recall, 
$f$
 is linear. Hence 
$$
 c_a =<l_a|A^{-1}w>+c_a<l_a|A^{-1}f_a> 
$$
 Thus we have seen, 
$$
 \fbox{$\displaystyle ( 1-<l_a|A^{-1}f_a> ) c_a = <l_a|A^{-1}w>  $}
$$
 and hence
$$
 \fbox{$\displaystyle 
 c_a = \frac{<l_a|A^{-1}w>}{ 1-<l_a|A^{-1}f_a> }\,, 
 \mbox{ when } 1-<l_a|A^{-1}f_a> \not= 0 \,. 
       $} 
$$  
 Finally 
$$
\fbox{ 
$\displaystyle
 (A-f_a<l_a|)^{-1}w
  = A^{-1}w+ A^{-1}f_a\frac{<l_a|A^{-1}w>}{ 1-<l_a|A^{-1}f_a> }\,, 
 \mbox{ when } 1-<l_a|A^{-1}f_a> \not= 0 \,. 
       $} 
$$

\newpage 

\subsubsection
{ Rank-Two Perturbations. }

 A problem connected with Rank-Two Perturbations looks quite similarly:
 We know  
$ A^{-1} $
 and need 
$$
 (A-f_a<l_a|-f_b<l_b|)^{-1}  
$$
 More precisely, we need to solve the associated equation.

 We can do it in two ways: we can iterate the final formulae 
 of the previous subsubsection,
 and we can iterate the argumentations of one. 
 In the former way, we first calculate 
$$
 (A-f_a<l_a|)^{-1} 
$$
 and then 
$$ 
 (B-f_b<l_b|)^{-1} \mbox{ where } B:= A-f_a<l_a| \,.
$$
 Describe the latter way. 

 The equation to be solved is this:
$$
 Av-f_a<l_a|v>-f_b<l_b|v>=w 
$$
 We rewrite it firstly as  
$$
 Av=w+f_a<l_a|v>+f_b<l_b|v>
$$
 and this relation, we rewrite it as 
$$
 v=A^{-1}(w + f_a<l_a|v> + f_b<l_b|v>) 
  = A^{-1}w+A^{-1}f_a<l_a|v>+A^{-1}f_b<l_b|v>
$$
 where we have used the fact that 
$A^{-1}$
 is linear.
 Thus,
$$
\fbox{ 
$\begin{array}{c}\displaystyle
 v=A^{-1}(w+c_a f_a+c_b f_b)
 = A^{-1}w+c_a A^{-1}f_a+c_b A^{-1}f_b \,, \\
\qquad\mbox{ where }  c_a:=<l_a|v>, c_b:=<l_b|v>
\end{array}$}
$$
 Apply now this form to itself and to the previous relationship. 
 Then obtain,  
$$
\begin{array}{ccc}
 \makebox[5ex][l]{$c_a A^{-1}f_a+c_b A^{-1}f_b$} &&\\
 &=& A^{-1}f_a<l_a|A^{-1}w + c_a A^{-1}f_a+ c_b A^{-1}f_b> \\
 & & { }  +
     A^{-1}f_b<l_b|A^{-1}w + c_a A^{-1}f_a+ c_b A^{-1}f_b>
\end{array}
$$
 To fulfil these relations, it is sufficient to satisfy the relations 
\footnote{
 if 
$ A^{-1}f_a, A^{-1}f_b $ 
 are linearly independent, then it is also necessary, 
 to do ones 
           } 

$$
\begin{array}{ccc}
 c_a &=& <l_a|A^{-1}w + c_a A^{-1}f_a+ c_b A^{-1}f_b> \\
 c_b &=& <l_b|A^{-1}w + c_a A^{-1}f_a+ c_b A^{-1}f_b>
\end{array}
$$
 i.e.,
\footnote{
 recall, 
$<l_a|$
 and 
$<l_b|$
 are linear
          }
$$
\fbox{
$\displaystyle
\Big(
\begin{array}{cc}
 1-<l_a|A^{-1}f_a> &  -<l_a|A^{-1}f_b>\\
  -<l_b|A^{-1}f_a> & 1-<l_b|A^{-1}f_b>\\
\end{array}
\Big)\Big(
\begin{array}{c}
 c_a \\ c_b
\end{array}
\Big)
=
\Big(
\begin{array}{c}
 <l_a|A^{-1}w> \\ <l_b|A^{-1}w> 
\end{array}
\Big)
$
}
$$
 Therefore, if 
$detrminant \not= 0$,  
$$
\Big(
\begin{array}{c}
 c_a \\ c_b
\end{array}
\Big)
 =
\Big(
\begin{array}{cc}
 1-<l_a|A^{-1}f_a> &  -<l_a|A^{-1}f_b>\\
  -<l_b|A^{-1}f_a> & 1-<l_b|A^{-1}f_b>\\
\end{array}
\Big)^{-1}
\Big(
\begin{array}{c}
 <l_a|A^{-1}w> \\ <l_b|A^{-1}w> 
\end{array}
\Big)
$$
$$
\Big(
\begin{array}{c}
 c_a \\ c_b
\end{array}
\Big)
 = 
 \frac{1}{determinant}
\Big(
\begin{array}{cc}
 1-<l_b|A^{-1}f_b> &  <l_a|A^{-1}f_b>\\
   <l_b|A^{-1}f_a> & 1-<l_a|A^{-1}f_a>\\
\end{array}
\Big)
\Big(
\begin{array}{c}
 <l_a|A^{-1}w> \\ <l_b|A^{-1}w> 
\end{array}
\Big)
$$
 Finally, 
$$
 ( A-f_a<l_a|-f_b<l_b| )^{-1}w = A^{-1}w + c_aA^{-1}f_a + c_bA^{-1}f_b
$$
$$
v=A^{-1}w 
 +(
\begin{array}{ccc}
 A^{-1}f_a \oplus A^{-1}f_b
\end{array} 
 )
\left(
\begin{array}[c]{cc}
 1-<l_a|A^{-1}f_a> &  -<l_a|A^{-1}f_b>\\
  -<l_b|A^{-1}f_a> & 1-<l_b|A^{-1}f_b>\\
\end{array}
\right)^{-1}
\left(
\begin{array}[c]{c}
 <l_a|A^{-1}w> \\ <l_b|A^{-1}w> 
\end{array}
\right)
$$

\newpage\subsection
{ Finite Rank Perturbations of Matrix-operator. }  

 The variant of the previous problems, we will deal, 
 looks like this: 
 Suppose, we can solve an equation array 
$$
\Big(
\begin{array}{cc}
 A_{11} & 0 \\
 0 &  A_{22} \\
\end{array}
\Big)
\Big(
\begin{array}{c}
 v_{01} \\ v_{02}
\end{array}
\Big)
 =
\Big(
\begin{array}{c}
 w_{01} \\ w_{02}
\end{array}
\Big)
$$
 and suppose, we need to solve an equation array   
$$
\Big(
\begin{array}{cc}
 A_{11}-f_{11}<l_1|_1  &        -f_{12}<l_2|_2 \\
       -f_{21}<l_1|_1  &  A_{22}-f_{22}<l_2|_2 \\
\end{array}
\Big)
\Big(
\begin{array}{c}
 v_1 \\ v_2
\end{array}
\Big)
 =
\Big(
\begin{array}{c}
 w_1 \\ w_2
\end{array}
\Big)
$$
 i.e. 
$$
\Big(
\begin{array}{cc}
 A_{11}v_1 -f_{11}<l_1|v_1> -f_{12}<l_2|v_2> \\
       -f_{21}<l_1|v_1>  +  A_{22}v_2-f_{22}<l_2|v_2> \\
\end{array}
\Big)
 =
\Big(
\begin{array}{c}
 w_1 \\ w_2
\end{array}
\Big) \,, 
$$
 where, of course,
$A_{11}, A_{22}$ 
 are two given invertible linear operators,  
 and 
$<l_1|, <l_2|$
 are two given linear functionals; the elements 
$f_{11}, f_{12}, f_{21}, f_{22}, w_1, w_2 $  
 are given, as well.

 One can show, it is exactly a rank-two case, an we can use 
 the {\em final formalae} of the previous subsection, but we prefer 
 direct reasoning 
 and we imitate the {\em arguing} of the previous subsection  
 and do as follows: 

 Firstly, write
$$
\Big(
\begin{array}{c}
 A_{11}v_1 \\ A_{22}v_2
\end{array}
\Big)
 = 
\Big(
\begin{array}{cc}
 w_1 + f_{11}<l_1|v_1> + f_{12}<l_2|v_2> \\
 w_2 + f_{21}<l_1|v_1> + f_{22}<l_2|v_2> \\
\end{array}
\Big)
$$
 and define
$$
\Big(
\begin{array}{c}
 c_1 \\ c_2
\end{array}
\Big)
 :=
\Big(
\begin{array}{c}
 <l_1|v_1> \\ <l_2|v_2> 
\end{array}
\Big)
$$
 Then write
$$
\Big(
\begin{array}{c}
 A_{11}v_1 \\ A_{22}v_2
\end{array}
\Big)
 = 
\Big(
\begin{array}{cc}
 w_1 + f_{11}c_1 + f_{12}c_2 \\
 w_2 + f_{21}c_1 + f_{22}c_2 \\
\end{array}
\Big)
$$
 Then 
$$
\Big(
\begin{array}{c}
 v_1 \\ v_2
\end{array}
\Big)
 = 
\Big(
\begin{array}{cc}
  A_{11}^{-1}w_1 + A_{11}^{-1}f_{11}c_1 + A_{11}^{-1}f_{12}c_2 \\
  A_{22}^{-1}w_2 + A_{22}^{-1}f_{21}c_1 + A_{22}^{-1}f_{22}c_2 \\
\end{array}
\Big)
$$
 Then 
$$
\Big(
\begin{array}{c}
 c_1 \\ c_2
\end{array}
\Big)
 = 
\Big(
\begin{array}{c}
 <l_1|v_1> \\ <l_2|v_2> 
\end{array}
\Big)
 = 
\Big(
\begin{array}{cc}
 <l_1| A_{11}^{-1}w_1 + A_{11}^{-1}f_{11}c_1 + A_{11}^{-1}f_{12}c_2> \\
 <l_2| A_{22}^{-1}w_2 + A_{22}^{-1}f_{21}c_1 + A_{22}^{-1}f_{22}c_2> \\
\end{array}
\Big)
$$
 Recall, 
$l_1, l_2$ 
 are linear. Therefore
$$
\Big(
\begin{array}{c}
 c_1 \\ c_2
\end{array}
\Big)
 = 
\Big(
\begin{array}{cc}
 <l_1| A_{11}^{-1}w_1> + <l_1|A_{11}^{-1}f_{11}>c_1 
                       + <l_1|A_{11}^{-1}f_{12}>c_2 \\
 <l_2| A_{22}^{-1}w_2> + <l_2|A_{22}^{-1}f_{21}>c_1 
                       + <l_2|A_{22}^{-1}f_{22}>c_2 \\
\end{array}
\Big)
$$
$$
\Big(
\begin{array}{cc}
 c_1 - <l_1|A_{11}^{-1}f_{11}>c_1 - <l_1|A_{11}^{-1}f_{12}>c_2 \\
     - <l_2|A_{22}^{-1}f_{21}>c_1 + c_2 - <l_2|A_{22}^{-1}f_{22}>c_2 \\
\end{array}
\Big)
 = 
\Big(
\begin{array}{cc}
 <l_1| A_{11}^{-1}w_1> \\
 <l_2| A_{22}^{-1}w_2> 
\end{array}
\Big)
$$
$$
\Big(
\begin{array}{cc}
 1 - <l_1|A_{11}^{-1}f_{11}>  &    - <l_1|A_{11}^{-1}f_{12}> \\
   - <l_2|A_{22}^{-1}f_{21}>  &  1 - <l_2|A_{22}^{-1}f_{22}> \\
\end{array}
\Big)
\Big(
\begin{array}{cc}
 c_1 \\
 c_2 \\
\end{array}
\Big)
 = 
\Big(
\begin{array}{cc}
 <l_1| A_{11}^{-1}w_1> \\
 <l_2| A_{22}^{-1}w_2> \\
\end{array}
\Big)
$$
 Thus we conclude 
$$
\Big(
\begin{array}{c}
 v_1 \\ v_2
\end{array}
\Big)
 = 
\Big(
\begin{array}{c}
 A_{11}^{-1}w_1 \\  A_{22}^{-1}w_2
\end{array}
\Big)
 +
\Big(
\begin{array}{cc}
 A_{11}^{-1}f_{11}  &  A_{11}^{-1}f_{12} \\
 A_{22}^{-1}f_{21}  &  A_{22}^{-1}f_{22} \\
\end{array}
\Big)
\Big(
\begin{array}{c}
 c_1 \\ c_2
\end{array}
\Big)
$$
$$
\Big(
\begin{array}{cc}
 c_1 \\
 c_2 \\
\end{array}
\Big)
 = 
\frac{1}{det}
\Big(
\begin{array}{cc}
 1 - <l_2|A_{22}^{-1}f_{22}> &      <l_1|A_{11}^{-1}f_{12}> \\
    <l_2|A_{22}^{-1}f_{21}>  &  1 - <l_1|A_{11}^{-1}f_{11}> \\
\end{array}
\Big)
\Big(
\begin{array}{cc}
 <l_1| A_{11}^{-1}w_1> \\
 <l_2| A_{22}^{-1}w_2> \\
\end{array}
\Big)
$$
 where 
$$
  det 
  := 
 (1 - <l_1|A_{11}^{-1}f_{11}>)(1 - <l_2|A_{22}^{-1}f_{22}>) 
 - <l_1|A_{11}^{-1}f_{12}><l_2|A_{22}^{-1}f_{21}>  
$$
 Of course, we need here 
$det \not= 0 $. 
 Note, we have obtained, that the values of 
$c_1, c_2$ must be such, as they are written. 
 But do they exist at all? Are the found 
$c_1, c_2$ 
 a solution of the problem? 
 The answer is: yes. 
 The way to verify it is plain: imitate the corresponding 
 part of the rank-one case,-- we omit details.

\bigskip 
 
 Finally we remark. 
 If  
$A_{11}$
 or 
$A_{22}$
 are not invertible, 
 but if 
$A_{11}-f_{11}<l_1|_1$ 
 or 
$A_{22}-f_{22}<l_2|_2$,
 if at least one of these operators 
 is invertible,  
 in this case 
 we can use one of 
 the following Frobenius formulae for the inverse of a block matrix: 
\begin{eqnarray*}
\makebox[7ex][l]{$
\left(\begin{array}{cc} A&B\\C&D \end{array}\right)^{-1}
  $}
 \\&=&
 \left(\begin{array}{cc}
 A^{-1}+A^{-1}B(D-CA^{-1}B)^{-1}C A^{-1}& - A^{-1}B(D-CA^{-1}B)^{-1}
 \\
 -(D-CA^{-1}B)^{-1} C A^{-1} & (D-CA^{-1}B)^{-1} 
 \\
 \end{array}\right)
\end{eqnarray*}
 or 
\begin{eqnarray*}
\makebox[7ex][l]{$
\left(\begin{array}{cc} A&B\\C&D \end{array}\right)^{-1}
  $}
 \\&=&
 \left(\begin{array}{cc}
 (A-BD^{-1}C)^{-1}  & -(A-BD^{-1}C)^{-1} BD^{-1}
 \\ 
 - D^{-1} C(A-BD^{-1}C)^{-1} & D^{-1}+ D^{-1} C(A-BD^{-1}C)^{-1} BD^{-1} 
 \end{array}\right) \,,
\end{eqnarray*}
 respectively. 

 We will not now concentrate ourselves upon details.

\newpage\section
{ Resolvents Formulae for the Models of Interaction} 

 Return to  

$$
 z
\left(\begin{array}{cc}
  q \\
  u
\end{array}\right)
 -
\left(\begin{array}{cc}
 -\Omega^2         
       &  \Omega^2<l|_2 \\
 4\gamma_c\delta_{\alpha,x_0}<1|_1 
       &  B -4\gamma_c\delta_{\alpha,x_0}<l|_2
\end{array}\right)
\left(\begin{array}{c}
 q \\
 u
\end{array}\right)
 = 
\left(\begin{array}{cc}
 w_1 \\
 w_2
\end{array}\right)
$$
 which we write also as 
$$
\left(\begin{array}{cc}
 z+\Omega^2         
       &  -\Omega^2<l|_2 \\
 -4\gamma_c\delta_{\alpha, x_0}<1|_1 
       &  z-B +4\gamma_c\delta_{\alpha, x_0}<l|_2
\end{array}\right)
\left(\begin{array}{c}
 q \\
 u
\end{array}\right)
 = 
\left(\begin{array}{cc}
 w_1 \\
 w_2
\end{array}\right)
$$
 and confer this latter expression with formulae of rank two perturbations. 
 We let 
$$
\left(\begin{array}{cc}
 A_{11}-f_{11}<l_1|_1  &        -f_{12}<l_2|_2 \\
       -f_{21}<l_1|_1  &  A_{22}-f_{22}<l_2|_2 \\
\end{array}\right)
 := 
\left(\begin{array}{cc}
 z+\Omega^2         
       &  -\Omega^2<l|_2 \\
 -4\gamma_c\delta_{\alpha, x_0}<1|_1 
       &  z-B +4\gamma_c\delta_{\alpha, x_0}<l|_2
\end{array}\right)
$$
 and next take  
$$
\begin{array}{cc}
  A_{11}  :=   z+\Omega^2   \,,\quad  <l_1|_1  := <1|_1 \,,
  &          
  A_{22}  :=  z-B   \,,\quad  <l_2|_2  := <l|_2
 \\
  f_{11}  := 0             &         f_{12} := \Omega^2 \\
  f_{21}  := 4\gamma_c\delta_{\alpha, x_0}  
                           &  f_{22} := -4\gamma_c\delta_{\alpha, x_0} \\
\end{array}
$$

 After substituting these values of 
$A_{...}$ 
 and 
$f_{...}$ 
 in 
$$
\left(\begin{array}{cc}
 1 - <l_1|A_{11}^{-1}f_{11}>  &    - <l_1|A_{11}^{-1}f_{12}> \\
   - <l_2|A_{22}^{-1}f_{21}>  &  1 - <l_2|A_{22}^{-1}f_{22}> \\
\end{array}\right)
\left(\begin{array}{cc}
 c_1 \\
 c_2 \\
\end{array}\right)
 = 
\left(\begin{array}{cc}
 <l_1| A_{11}^{-1}w_1> \\
 <l_2| A_{22}^{-1}w_2> \\
\end{array}\right)
$$
 we obtain 
$$
\begin{array}{rcl}
\makebox[45ex][l]{$\displaystyle 
\left(\begin{array}{cc}
 1 - <1|_1(z+\Omega^2)^{-1} 0>  
 &    - <1|_1(z+\Omega^2)^{-1} (\Omega^2)> \\
   -<l|_2(z-B)^{-1} (4\gamma_c\delta_{\alpha, x_0})>  
 &1-<l|_2(z-B)^{-1}(-4\gamma_c\delta_{\alpha, x_0})> \\
\end{array}\right)
\left(\begin{array}{cc}
 c_1 \\
 c_2 \\
\end{array}
\right)
  $}
\\[\medskipamount]\\&=& %
\left(\begin{array}{cc}
 <1|_1  (z+\Omega^2)^{-1}w_1> \\
 <l|_2 (z-B)^{-1}w_2> \\
\end{array}\right)
\end{array}
$$
  Then 
$$
\begin{array}{rcl}
\makebox[45ex][l]{$\displaystyle 
\left(\begin{array}{cc}
 1 
 &    - \dfrac{ \Omega^2 }{z+\Omega^2} \\[\medskipamount]\\
     -4\gamma_c<l|(z-B)^{-1}\delta_{\alpha, x_0}>  
 &  1+4\gamma_c<l|(z-B)^{-1}\delta_{\alpha, x_0}> \\
\end{array}\right)
\left(\begin{array}{cc}
 c_1 \\
 c_2 \\
\end{array}\right)
  $}
\\[\medskipamount]\\&=& %
\left(\begin{array}{cc}
 \dfrac{w_1}{ z+\Omega^2 } \\[\medskipamount]\\
 <l|(z-B)^{-1}w_2> \\
\end{array}\right)
\end{array}
$$


 Now let us choose 
$$
 B := c^2\Delta
$$
  In this case one can show that for any complex 
$z$
 such that 
$z \not\in (-\infty,0] $
$$
 (z-B)^{-1} \equiv  (z-c^2\Delta))^{-1}
$$
 exists 
 and is an integral operator which can be described as following
 (see e.g. Appendix A) : 
$$
 ((z-c^2\Delta))^{-1}F)(x) 
 = 
 \int_{-\infty}^{+\infty} \frac{i}{2k c^2}e^{ik|x-x'|} F(x')dx 
 \mbox{ where }  \quad k^2=-z/c^2, \quad Im\,k > 0 \\
$$

 If we choose in addition 
$$
 \delta_{\alpha, x_0}(x) := \delta(x-x_0)\,,\quad 
 <l|F> := F(x_0) 
$$
 then we obtain:  
$$
<l|(z-B)^{-1}\delta_{\alpha, x_0}> = \frac{i}{2k c^2} 
$$
 and equation 
$$
\begin{array}{rcl}
\makebox[45ex][l]{$\displaystyle 
\left(\begin{array}{cc}
 1 
 &    - \dfrac{ \Omega^2 }{z+\Omega^2} \\[\medskipamount]\\
     -4\gamma_c<l|(z-B)^{-1}\delta_{\alpha, x_0}>  
 &  1+4\gamma_c<l|(z-B)^{-1}\delta_{\alpha, x_0}> \\
\end{array}\right)
\left(\begin{array}{cc}
 c_1 \\
 c_2 \\
\end{array}\right)
  $}
\\[\medskipamount]\\&=& %
\left(\begin{array}{cc}
 \dfrac{w_1}{ z+\Omega^2 } \\[\medskipamount]\\
 <l|(z-B)^{-1}w_2> \\
\end{array}\right)
\end{array}
$$
 becomes as following:  
$$
\begin{array}{rcl}
\left(\begin{array}{cc}
 1                             &    - \dfrac{ \Omega^2 }{z+\Omega^2} 
 \\[\medskipamount]\\
  -4\gamma_c\dfrac{i}{2k c^2}  &  1+4\gamma_c\dfrac{i}{2k c^2} \\
\end{array}\right)
\left(\begin{array}{cc}
 c_1 \\[5ex]
 c_2 \\
\end{array}\right)
 &=& %
\left(\begin{array}{cc}
 \dfrac{w_1}{ z+\Omega^2 } \\[\medskipamount]\\
{\displaystyle 
 \int_{-\infty}^{+\infty} \frac{i}{2k c^2}e^{ik|x_0-x'|} w_2(x')dx } \\
\end{array}\right)
\end{array}
$$

 \addvspace{3\bigskipamount}

\begin{eqnarray*}
\left(\begin{array}{cc}
 c_1 \\[5ex]
 c_2 \\
\end{array}\right)
 &=&
\frac{1}{determinant}
\begin{array}{rcl}
\left(\begin{array}{cc}
 1+4\gamma_c\dfrac{i}{2k c^2} 
 &     \dfrac{ \Omega^2 }{z+\Omega^2} \\[\medskipamount]\\
    4\gamma_c\dfrac{i}{2k c^2}  
 &  1 \\
\end{array}\right)
\left(\begin{array}{cc}
 \dfrac{w_1}{ z+\Omega^2 } \\[\medskipamount]\\
{\displaystyle 
 \int_{-\infty}^{+\infty} \frac{i}{2k c^2}e^{ik|x_0-x'|} w_2(x')dx } \\
\end{array}\right)
\end{array}
 \\[\medskipamount]\\
 \\&&{}
 =
\frac{1}{determinant}
\left(\begin{array}{cc}
 (1+4\gamma_c\dfrac{i}{2k c^2}) \dfrac{w_1}{ z+\Omega^2 } 
 +  \dfrac{ \Omega^2 }{z+\Omega^2} 
    {\displaystyle 
     \int_{-\infty}^{+\infty} \frac{i}{2k c^2}e^{ik|x_0-x'|} w_2(x')dx }
\\[\medskipamount]\\
    4\gamma_c\dfrac{i}{2k c^2} \dfrac{w_1}{ z+\Omega^2 } 
  +
   {\displaystyle 
    \int_{-\infty}^{+\infty} \frac{i}{2k c^2}e^{ik|x_0-x'|} w_2(x')dx } \\
\end{array}\right)
\end{eqnarray*}
 where
$$
  k^2=-z/c^2, \quad Im\,k > 0 
$$
 and 
\begin{eqnarray*}
 determinant 
 &:=& 1+4\gamma_c\frac{i}{2k c^2} 
   - \frac{ \Omega^2 }{z+\Omega^2}4\gamma_c\frac{i}{2k c^2} 
 \\&&{ }
  =
  1+2\gamma_c\frac{i}{k c^2} 
   - \frac{ \Omega^2 }{z+\Omega^2}2\gamma_c\frac{i}{k c^2} 
\end{eqnarray*}
 i.e.,
\begin{eqnarray*}
 determinant 
&=&
 \frac{ k c^2(z+\Omega^2)+2\gamma_c i (z+\Omega^2) - 2\gamma_c i \Omega^2 } 
{k c^2(z+\Omega^2)} 
 =\frac{ k c^2(z+\Omega^2)+2\gamma_c i z }{k c^2(z+\Omega^2)} 
\\&=&\frac{ k c^2(-k^2c^2+\Omega^2)-2\gamma_c i k^2c^2 }
{k c^2(-k^2c^2+\Omega^2)} 
 =\frac{ -k^2c^2+\Omega^2 -2i\gamma_c k}
{ -k^2c^2+\Omega^2 } 
\\&=&-\frac{ k^2+\dfrac{2i\gamma_c}{c^2} k - \dfrac{\Omega^2}{c^2}}
{-k^2+\dfrac{\Omega^2}{c^2}} 
 =\frac{ (k+\dfrac{i\gamma_c}{c^2})^2
      - \dfrac{\Omega^2}{c^2} +(\dfrac{\gamma_c}{c^2})^2}
{k^2-\dfrac{\Omega^2}{c^2}} 
\end{eqnarray*}

 We observe that
$determinant$ 
 vanishes at two values of 
$k$:
$$
 k_{1,2} = -\dfrac{i\gamma_c}{c^2} 
      \pm  \sqrt{\dfrac{\Omega^2}{c^2} -(\dfrac{\gamma_c}{c^2})^2}
   \equiv  -\dfrac{i\gamma_c}{c^2} 
      \mp i\sqrt{- \dfrac{\Omega^2}{c^2} +(\dfrac{\gamma_c}{c^2})^2} \,.
$$                                
 But these values are forbidden to 
$k$
 because of the condition  
$Im\,k > 0 $ 
 and of course 
 because of the condition 
$\gamma_c > 0 $ 
  !!! 

 We temporarily defer discussing `what does it mean'. 
 In the next section, we will specially return to this question,  
 for the time being, we turn to calculating 
 the resolvent 
 associated with the Friedrichs model, i.e., with the equation array 
$$
 z
\left(\begin{array}{cc}
  q_0 \\
  \phi 
\end{array}\right)
 -
\left(\begin{array}{cc}
 -\Omega^2         
       &  \gamma_1<l|_2 \\
 4\gamma_{2,c}\delta_{\alpha,x_0}<1|_1 
       &  B 
\end{array}\right)
\left(\begin{array}{c}
 q_0 \\
 \phi 
\end{array}\right)
 = 
\left(\begin{array}{cc}
 w_1 \\
 w_2
\end{array}\right)
$$

 We take    
$$
\begin{array}{cc}
  A_{11}  :=   z+\Omega^2   \,,\quad  <l_1|_1  := <1|_1 \,,
  &          
  A_{22}  :=  z-B   \,,\quad  <l_2|_2  := <l|_2
 \\
  f_{11}  := 0             &         f_{12} := \gamma_1  \\
  f_{21}  := 4\gamma_{2,c}\delta_{\alpha,x_0}
                           &  f_{22} := 0     \\
\end{array}
$$
 and after substituting these values of 
$A_{...}$ 
 and 
$f_{...}$ 
 in 
$$
\left(\begin{array}{cc}
 1 - <l_1|A_{11}^{-1}f_{11}>  &    - <l_1|A_{11}^{-1}f_{12}> \\
   - <l_2|A_{22}^{-1}f_{21}>  &  1 - <l_2|A_{22}^{-1}f_{22}> \\
\end{array}\right)
\left(\begin{array}{cc}
 c_1 \\
 c_2 \\
\end{array}\right)
 = 
\left(\begin{array}{cc}
 <l_1| A_{11}^{-1}w_1> \\
 <l_2| A_{22}^{-1}w_2> \\
\end{array}\right)
$$
 we obtain 
$$
\begin{array}{rcl}
\makebox[45ex][l]{$\displaystyle 
\left(\begin{array}{cc}
 1 - <1|_1(z+\Omega^2)^{-1} 0> 
 &    - <1|_1(z+\Omega^2)^{-1} (\gamma_1)> \\
   -<l|_2(z-B)^{-1} (4\gamma_{2,c}\delta_{\alpha, x_0})> 
 & 1-<l|_2(z-B)^{-1}0> \\
\end{array}\right)
\left(\begin{array}{cc}
 c_1 \\
 c_2 \\
\end{array}\right)
  $}
\\[\medskipamount]\\&=& %
\left(\begin{array}{cc}
 <1|_1  (z+\Omega^2)^{-1}w_1> \\
 <l|_2 (z-B)^{-1}w_2> \\
\end{array}\right)
\end{array}
$$
  i.e., 
$$
\begin{array}{rcl}
\makebox[45ex][l]{$\displaystyle 
\left(\begin{array}{cc}
 1  
 &  -\dfrac{ \gamma_1 }{z+\Omega^2} 
 \\[\medskipamount]\\
       -4\gamma_{2,c}<l|(z-B)^{-1}\delta_{\alpha, x_0}>  
 &  1  \\
\end{array}\right)
\left(\begin{array}{cc}
 c_1 \\
 c_2 \\
\end{array}\right)
  $}
\\[\medskipamount]\\&=& %
\left(
\begin{array}{cc}
 \dfrac{w_1}{ z+\Omega^2 } \\[\medskipamount]\\
 <l|(z-B)^{-1}w_2> \\
\end{array}
\right)
\end{array}
$$

 Now let us take again  
$$
 B := c^2\Delta
$$
$$
 \delta_{\alpha, x_0}(x) := \delta(x-x_0)\,,\quad 
 <l|F> := F(x_0) 
$$
 Since in this case 
 for any complex 
$z$
 such that 
$z \not\in (-\infty,0] $
 the operator 
$$
 (z-B)^{-1} \equiv  (z-c^2\Delta))^{-1}
$$
 exists and since 
$$
 <l|(z-B)^{-1}\delta_{\alpha, x_0}> = \frac{i}{2k c^2} 
 \mbox{ where }  \quad k^2=-z/c^2, \quad Im\,k > 0 \\
$$
 the recent equation to 
$c_1, c_2$
 becomes as following:  
$$
\begin{array}{rcl}
\left(\begin{array}{cc}
 1 
 &    -\dfrac{ \gamma_1 }{z+\Omega^2} 
 \\[\medskipamount]\\
      -4\gamma_{2,c}\dfrac{i}{2k c^2} 
 &  1  \\
\end{array}\right)
\left(\begin{array}{cc}
 c_1 \\
 c_2 \\
\end{array}\right)
 &=& 
\left(\begin{array}{cc}
 \dfrac{w_1}{ z+\Omega^2 } \\[\medskipamount]\\
 <l|_2 (z-B)^{-1}w_2> \\
\end{array}\right)
\end{array}
$$

 \addvspace{3\bigskipamount}

\begin{eqnarray*}
\left(\begin{array}{cc}
 c_1 \\[5ex]
 c_2 \\
\end{array}\right)
 &=&
\frac{1}{determinant_F}
\left(\begin{array}{cc}
 1 
 &  \dfrac{ \gamma_1 }{z+\Omega^2} 
 \\[\medskipamount]\\
    4\gamma_{2,c}\dfrac{i}{2k c^2} 
 &  1  \\
\end{array}\right)
\left(\begin{array}{cc}
 \dfrac{w_1}{ z+\Omega^2 } \\[\medskipamount]\\
{\displaystyle 
 \int_{-\infty}^{+\infty} \frac{i}{2k c^2}e^{ik|x_0-x'|} w_2(x')dx } \\
\end{array}\right)
 \\[\medskipamount]\\
 \\&&{} 
 =
\frac{1}{determinant_F}
\left(\begin{array}{cc}
 \dfrac{w_1}{ z+\Omega^2 } 
 + \dfrac{ \gamma_1 }{z+\Omega^2} 
    {\displaystyle 
     \int_{-\infty}^{+\infty} \frac{i}{2k c^2}e^{ik|x_0-x'|} w_2(x')dx }
 \\[\medskipamount]\\
    4\gamma_{2,c}\dfrac{i}{2k c^2}\dfrac{w_1}{ z+\Omega^2 } 
  +
    {\displaystyle 
     \int_{-\infty}^{+\infty} \frac{i}{2k c^2}e^{ik|x_0-x'|} w_2(x')dx }   \\
\end{array}\right)
\end{eqnarray*}
 where
$$
  k^2=-z/c^2, \quad Im\,k > 0 
$$
 and 
\begin{eqnarray*}
 determinant_F 
 &:=& 1+\dfrac{ \gamma_1 }{z+\Omega^2}4\gamma_{2,c}\frac{i}{2k c^2} 
 \\&&{ }
  = 1+\dfrac{ \gamma_1 }{-k^2 c^2+\Omega^2}\gamma_{2,c}\frac{i}{k c^2}
 \\&&{ }
 \frac{(-k^2 c^2+\Omega^2)k c^2 
        +i\gamma_1\gamma_{2,c}}{(-k^2 c^2+\Omega^2)k c^2}
\end{eqnarray*}
 i.e.,
\begin{eqnarray*}
 determinant_F 
 &=&  
 \frac{(-k^2 +\Omega^2/c^2)k  
        +i\gamma_1\gamma_{2,c}/c^4}{(-k^2 +\Omega^2/c^2)k }
\end{eqnarray*}
 
\newpage 
 We conclude this section with two remarks. 
\begin{Remark}{ 1. }
 We have considered the operator 
$$
\left(\begin{array}{cc}
 z+\Omega^2         
       &  -\Omega^2<l|_2 \\
 -4\gamma_c\delta_{\alpha, x_0}<1|_1 
       &  z-B +4\gamma_c\delta_{\alpha, x_0}<l|_2
\end{array}\right)
$$
 as a perturbation of the operator 
$$
\left(\begin{array}{cc}
 z+\Omega^2  &  0\\
      0      &  z-B 
\end{array}\right) \,.
$$
 The perturbation, i.e. the operator  
$$
\left(\begin{array}{cc}
 0    &  -\Omega^2<l|_2 \\
 -4\gamma_c\delta_{\alpha,x_0}<1|_1 
       &  4\gamma_c\delta_{\alpha,x_0}<l|_2
\end{array}\right) 
$$
 is a {\bf rank-two} operator, we have already mentioned it. 
 If we initially take 
$$
\left(\begin{array}{cc}
 z   &  0\\
 0   &  z-B 
\end{array}\right)
$$
 as an unperturbated operator, then the perturbation becomes 
$$
\left(\begin{array}{cc}
 \Omega^2                             &  -\Omega^2<l|_2 \\
 -4\gamma_c\delta_{\alpha,x_0}<1|_1   &  4\gamma_c\delta_{\alpha,x_0}<l|_2
\end{array}\right) \,, 
$$
 and this operator is a {\bf rank-one} one. 

 Actually,
\begin{eqnarray*}
\left(\begin{array}{cc}
 \Omega^2                             &  -\Omega^2<l|_2 \\
 -4\gamma_c\delta_{\alpha,x_0}<1|_1   &  4\gamma_c\delta_{\alpha,x_0}<l|_2
\end{array}\right)  
 &=&
\Biggl|\begin{array}{cc}
 \Omega^2 \\ -4\gamma_c\delta_{\alpha,x_0}
\end{array}\Biggr> 
 <1 \oplus -l |
 \\
 &=&
\left(\begin{array}{cc}
 \Omega^2 \\ -4\gamma_c\delta_{\alpha,x_0}
\end{array}\right) 
\left(\begin{array}{ccc}
 <1|_1 &\oplus & -<l|_2
\end{array}\right) \,. 
\end{eqnarray*}
 We have used here the fact that, as an operator 
${\bf C}\to {\bf C}$, 
$ -\Omega^2<1|_1$ 
 coincides with the operator of multiplication by 
$ -\Omega^2$ :   
$$
 -\Omega^2<1|_1q> = -\Omega^2 q \,,\, q\in {\bf C} 
$$
 ---in other words, 
$$
 -\Omega^2<1|_1 =-\Omega^2 \cdot = -\Omega^2 I_{\bf C} = \Omega^2 \,. 
$$
 Of course,  we have identified 
 the operator multiplication by a number with  this number itself: 

 the operator multiplication by a number 
$\equiv $  this number itself. 

\end{Remark}

\newpage 
\begin{Remark}{ 2. }
 As for an explicit and detailed expression of 
$$
\left(\begin{array}{cc}
 z+\Omega^2         
       &  -\Omega^2<l|_2 \\
 -4\gamma_c\delta_{\alpha, x_0}<1|_1 
       &  z-B +4\gamma_c\delta_{\alpha, x_0}<l|_2
\end{array}\right)^{-1} \,, 
$$
 we have 
\begin{eqnarray*}
\Big(
\begin{array}{c}
 q \\ 
 u
\end{array}
\Big)
 &=& 
\Big(
\begin{array}{c}
 A_{11}^{-1}w_1 \\  A_{22}^{-1}w_2
\end{array}
\Big)
 +
\Big(
\begin{array}{cc}
 A_{11}^{-1}f_{11}  &  A_{11}^{-1}f_{12} \\
 A_{22}^{-1}f_{21}  &  A_{22}^{-1}f_{22} \\
\end{array}
\Big)
\Big(
\begin{array}{c}
 c_1 \\ 
 c_2
\end{array}
\Big)
\\[\smallskipamount]\\ &=& 
\Big(
\begin{array}{c}
 (z+\Omega^2)^{-1}w_1 \\  (z-B)^{-1}w_2
\end{array}
\Big)
 +
\Big(
\begin{array}{cc}
 (z+\Omega^2)^{-1}\cdot 0  &   (z+\Omega^2)^{-1}\Omega^2 \\
 (z-B)^{-1}4\gamma_c\delta_{\alpha, x_0} 
                           &  -(z-B)^{-1}4\gamma_c\delta_{\alpha, x_0} \\
\end{array}
\Big)
\Big(
\begin{array}{c}
 c_1 \\ 
 c_2
\end{array}
\Big)
\\[\smallskipamount]\\ &=& 
\Big(
\begin{array}{c}
 (z+\Omega^2)^{-1}w_1 \\  (z-B)^{-1}w_2
\end{array}
\Big)
 +
\Big(
\begin{array}{cc}
          0                &   (z+\Omega^2)^{-1}\Omega^2 \\
 (z-B)^{-1}4\gamma_c\delta_{\alpha, x_0} 
                           &  -(z-B)^{-1}4\gamma_c\delta_{\alpha, x_0} \\
\end{array} 
\Big)
\Big(
\begin{array}{c}
 c_1 \\ 
 c_2
\end{array}
\Big) \,,
\end{eqnarray*}
\begin{eqnarray*}
\makebox[4ex][l]{$\displaystyle 
\left(\begin{array}{cc}
 c_1 \\
 c_2 \\
\end{array}\right)\cdot det 
  $}
\\&=& 
\left(\begin{array}{cc}
 \Bigl(1+4\gamma_c<l|(z-B)^{-1}\delta_{\alpha, x_0}>\Bigr) 
 &     \dfrac{ \Omega^2 }{z+\Omega^2} 
\\[\medskipamount]\\
     4\gamma_c<l|(z-B)^{-1}\delta_{\alpha, x_0}>  
 &  1 \\
\end{array}\right)
\left(\begin{array}{cc}
 \dfrac{w_1}{ z+\Omega^2 } 
\\[\medskipamount]\\
 <l|(z-B)^{-1}w_2> \\
\end{array}\right)
\\[\medskipamount]\\ &=& 
\left(\begin{array}{cc}
 \Bigl(1+4\gamma_c<l|(z-B)^{-1}\delta_{\alpha, x_0}>\Bigr)
 \dfrac{w_1}{ z+\Omega^2 } 
 + \dfrac{ \Omega^2 }{z+\Omega^2}<l|(z-B)^{-1}w_2>
\\[\medskipamount]\\
 4\gamma_c<l|(z-B)^{-1}\delta_{\alpha, x_0}>\dfrac{w_1}{ z+\Omega^2 }  
 + <l|(z-B)^{-1}w_2> \\
\end{array}\right) \,, 
\end{eqnarray*} 
 where   
$$
 det 
 =
 1+4\gamma_c<l|(z-B)^{-1}\delta_{\alpha, x_0}> 
 -
 \frac{ \Omega^2 }{z+\Omega^2}
 4\gamma_c<l|(z-B)^{-1}\delta_{\alpha, x_0}>
 \,.
$$
 Notice, by the way, 
\begin{eqnarray*}
 det\cdot (z+\Omega^2) 
 &=&
 (z+\Omega^2)+4\gamma_c z <l|(z-B)^{-1}\delta_{\alpha, x_0}> 
 \\&&{}
 = \Bigl(1+4\gamma_c<l|(z-B)^{-1}\delta_{\alpha, x_0}>\Bigr)z +\Omega^2
 \,,
\end{eqnarray*}
 \begin{eqnarray*}
\makebox[4ex][l]{$\displaystyle 
\left(\begin{array}{cc}
 c_1 \\
 c_2 \\
\end{array}\right) det\cdot (z+\Omega^2)
  $}
\\[\medskipamount]\\ &=& 
\left(\begin{array}{cc}
 \Bigl(1+4\gamma_c<l|(z-B)^{-1}\delta_{\alpha, x_0}>\Bigr){w_1}
 +  \Omega^2 <l|(z-B)^{-1}w_2>
\\[\medskipamount]\\
 4\gamma_c<l|(z-B)^{-1}\delta_{\alpha, x_0}>{w_1}
 + (z+\Omega^2)<l|(z-B)^{-1}w_2> \\
\end{array}\right) \,. 
\end{eqnarray*} 

\end{Remark}

\newpage\section
{ Resonances, Resolvents, and the Second Sheet } 

 In the previous section 
 we have observed that the 
$determinant$ 
 vanishes at two values of 
$k$:
\begin{eqnarray*}
 k_{1,2} 
 &=& -\dfrac{i\gamma_c}{c^2} 
      \pm \sqrt{\dfrac{\Omega^2}{c^2} -(\dfrac{\gamma_c}{c^2})^2} \,.
 \\&&{} 
 \equiv
    -\dfrac{i\gamma_c}{c^2} 
     \mp i \sqrt{ -\dfrac{\Omega^2}{c^2} +(\dfrac{\gamma_c}{c^2})^2} \,.
\end{eqnarray*}
 The circumstance having excited an interest is that 
 these values are forbidden (!!!) to 
$k$
 because of the condition  
$Im\,k > 0 $ 
\footnote{ and of course 
 because of the condition 
$\gamma_c > 0 $} . 
 In this section, we return specially to this fact,  
 and we are now discussing `what does it mean'. 

 Firstly, we fix on 
$$
 det_{-1,+}(k)
  := 
 ((z+\Omega^2)determinant)^{-1}
 \equiv 
 ((-k^2 c^2+\Omega^2)determinant)^{-1}
$$
 as a function of 
$k$,
 ---we need rather this latter quantity than the  
$determinant$ .
 We have:
\begin{eqnarray*}
 det_{-1}(k,+)
&=&
 -\frac{1/c^2}
{ k^2+\dfrac{2i\gamma_c}{c^2} k - \dfrac{\Omega^2}{c^2}}
 =
  \frac{1/c^2}
{ (k+\dfrac{i\gamma_c}{c^2})^2 
    - \dfrac{\Omega^2}{c^2} +(\dfrac{\gamma_c}{c^2})^2} \,,
\\&&\qquad\qquad\qquad
 \quad (\mbox{ where } k^2=-z/c^2 \,,\, Im\,k > 0) \,.\\
\end{eqnarray*}
 The first factor which calls attention to itself is that 
 the written form of 
$det_{-1}(k)$
 in itself 
 need not the relation 
$k^2=-z/c^2 \,,\, Im\,k > 0$
 being fulfilled, and provokes to introduce the formal extension
 of 
$det_{-1}(k,+)$ , 
$Det_{-1}(k)$ 
 say, defined by   
\begin{eqnarray*}
 Det_{-1}(k) 
&:=&
 -\frac{1/c^2}
{ k^2+\dfrac{2i\gamma_c}{c^2} k - \dfrac{\Omega^2}{c^2}}
 \,,\,\qquad  
 k^2+\dfrac{2i\gamma_c}{c^2} k - \dfrac{\Omega^2}{c^2} \not=0 \,.
\end{eqnarray*}
 This extension 
$Det_{-1}(k)$ 
 is an analytic function of 
$k$, 
 and  
 at {\bf real} 
$k$ 
\begin{eqnarray*}
 |Det_{-1}(k)|^2 
 =|det_{-1,+}(k+i0))|^2 
&=& 
 \frac{1/c^4}
{ (k^2- \dfrac{\Omega^2}{c^2})^2+(\dfrac{2\gamma_c}{c^2} k)^2 } \,,
 \quad (k\in {\bf R}) \,.
\\
\end{eqnarray*}
 For a moment, let us renormalise 
$\Omega$
 and 
$\gamma_c$ 
 so that 
$$
 \frac{\Omega}{c} \longrightarrow \Omega \,,\,
 \frac{\gamma_c}{c^2} \longrightarrow \gamma \,,\, 
$$
 or choose the measure units so that 
$c=1$. 
 Then we have: 
\begin{eqnarray*}
 det_{-1,+}(k)
&=&
 -\frac{1}
{ k^2+2i\gamma k - \Omega^2}
 =
  \frac{1}
{ (k+i\gamma)^2 - \Omega^2 +\gamma^2} \,,
\\&&\qquad\qquad\qquad
 \quad (\mbox{ where } k^2=-z \,,\, Im\,k > 0) \,,\\
\end{eqnarray*}
\begin{eqnarray*}
 |det_{-1}(k+i0))|^2 
&=& 
 \frac{1}
{ (k^2- \Omega^2)^2+(2\gamma k)^2 } \,,
 \quad (k\in {\bf R}) \,.
\\
\end{eqnarray*}
\newpage 
 We had already seen a similar expression. 
 We had seen it in the previous paper. 

\addvspace{\bigskipamount} 

 In the previous paper
\footnote{we mean [Ch03-1]},
 in subsection 1.4, where just 
$c=1$, 
 we had seen:
 given an incident wave of the kind 
$$
  A_{mp}\sin(k(x+t)) \quad 
 \mbox{ where } 
 \quad k\in {\bf R} \,, 
 \quad A_{mp}\in {\bf R} \,, 
$$
 then 
\begin{eqnarray*}
 q(t)
 &=& 
 \frac{\Omega^2 A_{mp}}{ (-k^2+\Omega^2)^2 + (2\gamma k)^2 } 
  \left( 
    ( -k^2+\Omega^2 )\sin(kt) -2\gamma k \cos(kt)
  \right) + const_1
\\&&{} 
 + const_1 
 + 
 \frac{2\gamma}{\Omega^2}\frac{\partial q(t)}{\partial t}\Big|_{t=0}
 +
   e^{-\gamma t} 
  \Bigl(const_1\cos(\Omega_{\gamma} t) 
        +const_2 \sin(\Omega_{\gamma} t)
  \Bigr) \,;
\end{eqnarray*}
 we can write this relation as follows: 
\begin{eqnarray*}
 q(t)
 &=& 
 \frac{\Omega^2 A_{mp}}{\sqrt{ (-k^2+\Omega^2)^2 + (2\gamma k)^2}} 
   \sin(kt +\phi_k )
\\&&{} 
 + const_1
 +
 \frac{2\gamma}{\Omega^2}\frac{\partial q(t)}{\partial t}\Big|_{t=0}
 +
   e^{-\gamma t} 
  \Bigl(const_1\cos(\Omega_{\gamma} t) 
        +const_2 \sin(\Omega_{\gamma} t)
  \Bigr)
\end{eqnarray*}
 where
$\phi_k$ 
 is such that 
$$
  \cos\phi_k = \frac{-k^2+\Omega^2}{\sqrt{(-k^2+\Omega^2)^2 + (2\gamma k)^2 }} 
\,,\,
  \sin\phi_k = -\frac{2\gamma k}{\sqrt{(-k^2+\Omega^2)^2 + (2\gamma k)^2 }} 
\,. 
$$
 We had also seen that 
 the field 
$u(t,x)$ 
 is given by 
\begin{eqnarray*}
 u(t,x) 
 &=&
 \left\{ 
 \begin{array}{ccl}
 \displaystyle
 Q(t-|x|)-A_{mp}\sin(k(t-|x|))          &,& \mbox{ if } 0 \leq t-|x| \\ 
                0                &,& \mbox{ if } t-|x| < 0 \leq t \\ 
 \end{array}
 \right\} 
 + A_{mp}\sin(k(x+t)) \,, 
\end{eqnarray*}
 where 
\begin{eqnarray*}
 Q(t)-A_{mp}\sin(kt)
 &=& 
 \frac{-2\gamma A_{mp}k}{ (-k^2+\Omega^2)^2 + (2\gamma k)^2 } 
  \left( 
     2\gamma k\sin(kt) + ( -k^2+\Omega^2 ) \cos(kt)
  \right) 
 \\&&{}
 +
 \frac{2\gamma}{\Omega^2}\frac{\partial q(t)}{\partial t}\Big|_{t=0}
  +e^{-\gamma t} 
  \Bigl(const_3\cos(\Omega_{\gamma} t) 
        +const_4 \sin(\Omega_{\gamma} t)
  \Bigr)
\end{eqnarray*}
 and, thus, at points of the left real half-line 
 the field is given by the relationships: 
\begin{eqnarray*}
 u(t,x) 
 &=&
 \left\{ 
 \begin{array}{ccl}
 \displaystyle
 Q(t+x)    &,& \mbox{ if } x < 0, 0 \leq t+x \\ 
     0     &,& \mbox{ if } x < 0, t+x < 0 \leq t \\ 
 \end{array}
 \right\} 
\end{eqnarray*}

\begin{eqnarray*}
\makebox[4ex][l]{$\displaystyle 
 Q(t) 
 - 
 \frac{2\gamma}{\Omega^2}\frac{\partial q(t)}{\partial t}\Big|_{t=0}
  -e^{-\gamma t} 
  \Bigl(const_3\cos(\Omega_{\gamma} t) 
        +const_4 \sin(\Omega_{\gamma} t)
  \Bigr)
 $}
 \\&=& 
 \left( 
 \frac{-(2\gamma k)^2 A_{mp}}{ (-k^2+\Omega^2)^2 + (2\gamma k)^2 } 
  + A \right)\sin(kt)
  +
 \frac{-2\gamma A_{mp}k( -k^2+\Omega^2 )}{ (-k^2+\Omega^2)^2 + (2\gamma k)^2 } 
 \cos(kt)
 \\&&{} 
 = 
 \frac{(-k^2+\Omega^2)^2 A_{mp}}{ (-k^2+\Omega^2)^2 + (2\gamma k)^2 } 
 \sin(kt)
  +
 \frac{-2\gamma A_{mp}k( -k^2+\Omega^2 )}{ (-k^2+\Omega^2)^2 + (2\gamma k)^2 } 
 \cos(kt)
 \\&&{}\quad 
 = 
 \frac{(-k^2+\Omega^2)A_{mp}}{ (-k^2+\Omega^2)^2 + (2\gamma k)^2 } 
 \left( 
 (-k^2+\Omega^2)
 \sin(k(t))
  -
 2\gamma k
 \cos(kt)
 \right) 
 \\&&{}\qquad 
 = 
 \frac{(-k^2+\Omega^2)A_{mp}}{\sqrt{ (-k^2+\Omega^2)^2 + (2\gamma k)^2 }} 
 \sin(kt+\phi_k)
 \,.
\end{eqnarray*}

 In addition, 
\begin{eqnarray*}
\makebox[4ex][l]{$\displaystyle 
 q(t) - Q(t) 
  -e^{-\gamma t} 
  \Bigl((const_1-const_3)\cos(\Omega_{\gamma} t) 
        +(const_2-const_4)\sin(\Omega_{\gamma} t)
  \Bigr)
 $}
 \\&=& 
 \frac{k^2 A_{mp}}{ (-k^2+\Omega^2)^2 + (2\gamma k)^2 } 
 \left( 
 (-k^2+\Omega^2)
 \sin(k(t))
  -
 2\gamma k
 \cos(kt)
 \right) 
 \\&&{}\qquad 
 = 
 \frac{k^2 A_{mp}}{\sqrt{ (-k^2+\Omega^2)^2 + (2\gamma k)^2 }} 
 \sin(kt+\phi_k)
 \,.
\end{eqnarray*}

 Now, let us concentrate ourselves upon 
 the expressions of the amplitudes 
 of the harmonic parts of 
$q(t)$ ,  
$Q(t)$ 
 and 
$q(t)-Q(t)$ , 
 ---expressions as functions of 
$k$ ,---
 i.e., 
 let us concentrate upon oscillator amplitude, 
 transmitted wave one, and a `deformation' amplitude. 
 These amplitudes are exactly   
\begin{eqnarray*}
 amplitude_q(k) 
 &=&
 \frac{\Omega^2 A_{mp}}{\sqrt{ (-k^2+\Omega^2)^2 + (2\gamma k)^2}}\,, 
 \\
 amplitude_Q(k) 
 &=&
 \frac{(-k^2+\Omega^2)A_{mp}}{\sqrt{ (-k^2+\Omega^2)^2 + (2\gamma k)^2}}\,, 
 \\
 amplitude_{qQ}(k) 
 &=&
 \frac{k^2 A_{mp}}{\sqrt{ (-k^2+\Omega^2)^2 + (2\gamma k)^2 }}\,; 
\end{eqnarray*}
 we will also need 
$ amplitude_q(k)^2 $ , 
$ amplitude_Q(k)^2 $ ,
 and 
$ amplitude_{qQ}(k)^2 $ : 
$$
 amplitude_q(k)^2 
 =
 \frac{\Omega^4 A_{mp}^2}{{ (-k^2+\Omega^2)^2 + (2\gamma k)^2}}\,, 
$$
$$
 amplitude_Q(k)^2 
 =
 \frac{(-k^2+\Omega^2)^2 A_{mp}^2}{(-k^2+\Omega^2)^2 + (2\gamma k)^2}\,, 
$$
$$
 amplitude_{qQ}(k)^2 
 =
 \frac{(k^2)^2 A_{mp}^2}{(-k^2+\Omega^2)^2 + (2\gamma k)^2}\,. 
$$
 After standard transformations--- 
$$
 (-k^2+\Omega^2)^2 + (2\gamma k)^2
 =
 k^4 -2(\Omega^2-2\gamma^2)k^2+\Omega^4
 =
 (k^2-(\Omega^2-2\gamma^2))^2 + 4\gamma^2(\Omega^2-\gamma^2)
$$
 ---we also write
$ amplitude_q(k)^2 $ 
 and 
$ amplitude_{qQ}(k)^{2} $ 
 as follows: 
$$
 amplitude_q(k)^2 
 =
 \frac{\Omega^4 A_{mp}^2}{{ (-k^2+\Omega^2)^2 + (2\gamma k)^2}} 
 =
 \frac{\Omega^4 A_{mp}^2}
{(k^2-(\Omega^2-2\gamma^2))^2 + 4\gamma^2(\Omega^2-\gamma^2)} \,,
$$
$$
 amplitude_{qQ}(k)^{2} 
 =
 \frac{\Omega^4 A_{mp}^2}{1 -2(\Omega^2-2\gamma^2)k^{-2}+\Omega^4 k^{-4}} \,.
$$
\footnote{
 In addition, we have 
$$
 amplitude_Q(k)^2 
 =
 \frac{A_{mp}^2}{1 - (2\gamma)^2/(-k^2+\Omega^2)
 + (2\gamma)^2\Omega^2/(-k^2+\Omega^2)^2 }\,. 
$$
 } 

 We emphasise: in transforming no special assumption has been made. 
 But now recall, we have been discussing 
 incident wave 
 where 
$k$
 is a wave-number, 
 so, we presume that 
$k\in {\bf R}$ . 

\addvspace{\medskipamount}

 Thus, since 
$k\in {\bf R}$ , 
 we had seen:
 there is a value of a system parameter, 
 ---this parameter is 
$k$, 
 the wave-number of the incident wave, 
 ---such that 
 the oscillator amplitude becomes {\bf maximal}. 
 This phenomenon is called a {\bf resonance}, 
 in the proper, usual sense of the word.  
 Temporarily, we call this phenomenon {\bf resonance of the first kind}. 

\addvspace{\medskipamount}

 We had also seen:
 there is a value of a system parameter, 
 ---this parameter is 
$k$ again, 
 the wave-number of the incident wave, 
 ---such that 
 the incident harmonic wave is {\bf completely reflected}. 
 This phenomenon is a kind of {\bf resonance}, 
 we will temporarily call it {\bf resonance of the second kind}.   

\addvspace{\medskipamount}
 
 We have finally seen: 
 there is a value of 
$k$ 
 such that 
 amplitude of 
$q(t)-Q(t)$
 becomes maximal. 
 This phenomenon is also a kind of {\bf resonance}, 
 we will temporarily call it {\bf resonance of the third kind}.   
 
\addvspace{\medskipamount}

 Currently, the first kind resonant values of 
$k$ 
 are exactly 
$$
 k_{1st.res} = \pm\sqrt{\Omega^2 - 2\gamma^2} 
$$
 the second kind ones are 
$$
 k_{2nd.res} = \pm\Omega 
$$
 and the third kind resonant values of 
$k$ 
 are 
$$
 k_{3rd.res} = \pm\sqrt{\frac{\Omega^4}{\Omega^2 - 2\gamma^2}} \,.
$$

\addvspace{\bigskipamount}

 Another standard of transforming the denominator 
 in the expressions of 
$amplitude_q(k)^2$ 
 and  
$amplitude_Q(k)^2$ 
 is this:
$$
 (-k^2+\Omega^2)^2 + (2\gamma k)^2 
 =
 (k^2 + 2i\gamma k -\Omega^2 )
 (k^2 - 2i\gamma k -\Omega^2 ) \,.
$$
 Thus, we observe that 
$$
 amplitude_q(k)^2 
 =
 \frac{\Omega^4 A_{mp}^2}{{ (-k^2+\Omega^2)^2 + (2\gamma k)^2}} 
 =
 \frac{\Omega^4 A_{mp}^2}
{ (k^2 + 2i\gamma k -\Omega^2 )
  (k^2 - 2i\gamma k -\Omega^2 )}
$$
$$
 amplitude_Q(k)^2 
 =
 \frac{(-k^2+\Omega^2)^2 A_{mp}^2}{(-k^2+\Omega^2)^2 + (2\gamma k)^2}
 =
 \frac{(-k^2+\Omega^2)^2 A_{mp}^2}
{ (k^2 + 2i\gamma k -\Omega^2 )
  (k^2 - 2i\gamma k -\Omega^2 )}
$$
$$
 amplitude_{qQ}(k)^2 
 =
 \frac{(k^2)^2 A_{mp}^2}{(-k^2+\Omega^2)^2 + (2\gamma k)^2}
 =
 \frac{(k^2)^2 A_{mp}^2}
{ (k^2 + 2i\gamma k -\Omega^2 )
  (k^2 - 2i\gamma k -\Omega^2 )}
$$
 and recall that  
\begin{eqnarray*}
 det_{-1,+}(k)
&=&
 -\frac{1}
{ k^2+2i\gamma k - \Omega^2}
 =
  \frac{1}
{ (k+i\gamma)^2 - \Omega^2 +\gamma^2} \,,
\\&&\qquad\qquad\qquad
 \quad (\mbox{ where } k^2=-z \,,\, Im\,k > 0) \,,\\
\end{eqnarray*}
\begin{eqnarray*}
 |det_{-1}(k+i0))|^2 
&=& 
 \frac{1}
{ (k^2- \Omega^2)^2+(2\gamma k)^2 } 
 =
 \frac{1}
{ (k^2 + 2i\gamma k -\Omega^2 ) 
  (k^2 - 2i\gamma k -\Omega^2 )} 
 \quad (k\in {\bf R}) \,.
\\
\end{eqnarray*}
 First we focus on expressions of 
$amplitude_q(k)^2$
 and 
$det_{-1}(k)$ .
 In the first moment we tend to write
$$
 amplitude_q(k)^2 ={\Omega^4 A_{mp}^2}det_{-1,+}(k)det_{-1,+}(-k) \,.
$$
 We may not do it. 
 There are at least two reasons, both of them are connected with the domains. 

 If 
$Im\,k > 0$
 then
$Im\,(-k) < 0$. 
 Hence, if 
$det_{-1,+}(k)$
 is defined, then 
$det_{-1,+}(-k)$
 is not. 

 Secondly, if we deal with 
$det_{-1,+}(k)$ , 
 then    
$ Im\,k > 0$ ; 
 if we deal with 
$amplitude_q(k)^2$
 then    
$ k\in {\bf R}$ ,
 i.e.,
$Im\,k = 0$; 
 so, we must deal here with two `different' 
$k$-s. 

 It is no surprise: 
$k$
 as a parameter of 
$amplitude_q(k)^2$
 means wave-number of incident wave, whereas 
$k$
 as a parameter of 
$det_{-1}(k,+)$ 
 is a term in describing of a resolvent of an operator. 

 In other words,
$k$
 in 
$amplitude_q(k)^2$
 and 
$k$
 in 
$det_{-1}(k,+)$ 
 are of different natures.

 Nevertheless, the relation 
$$
 amplitude_q(k)^2={\Omega^4 A_{mp}^2}|det_{-1,+}(k+i0))|^2
$$
 is mathematically correct, and we see: 
 in order to find resonant values of the amplitude 
 we have {\bf to find positions of extremal values of 
 a function, which is absolute values function 
 of boundery values function 
 of an analytic function, analytic in the open upper half-plane
 ($Im\,k >0$) }.

 \addvspace{\bigskipamount}

 Let us discuss 
 the recent idea in terms of analytic functions.   
 We see, the written form of 
$amplitude_q(k)^2 $, $amplitude_Q(k)^2 $
 and 
 $amplitude_{qQ}(k)^2 $ 
 in itself 
 need not the relation 
$k^2=-z/c^2 \,,\, Im\,k > 0$
 being fulfilled, and provokes to introduce the formal extensions
 of these quantities, 
$Amplitude_q(k)^2 $, $Amplitude_Q(k)^2 $
 and 
 $Amplitude_{qQ}(k)^2 $ 
 say, defined by   
$$
 Amplitude_q(k)^2 
 :=
 \frac{\Omega^4 A_{mp}^2}{{ (-k^2+\Omega^2)^2 + (2\gamma k)^2}} 
 =
 \frac{\Omega^4 A_{mp}^2}{ (k^2 + 2i\gamma k -\Omega^2 ) 
                      (k^2 - 2i\gamma k -\Omega^2 )} 
$$
$$
 Amplitude_Q(k)^2 
 :=
 \frac{(-k^2+\Omega^2)^2 A_{mp}^2}{(-k^2+\Omega^2)^2 + (2\gamma k)^2}
 =
 \frac{(-k^2+\Omega^2)^2 A_{mp}^2}{ (k^2 + 2i\gamma k -\Omega^2 ) 
                      (k^2 - 2i\gamma k -\Omega^2 )} 
$$
$$
 Amplitude_{qQ}(k)^2 
 :=
 \frac{(k^2)^2 A_{mp}^2}{(-k^2+\Omega^2)^2 + (2\gamma k)^2}
 =
 \frac{(k^2)^2 A_{mp}^2}{ (k^2 + 2i\gamma k -\Omega^2 ) 
                      (k^2 - 2i\gamma k -\Omega^2 )} 
$$
 wherever 
$$
 (-k^2+\Omega^2)^2 + (2\gamma k)^2 \not=0\,.
$$

 These extensions 
 are analytic functions of 
$k$, with poles at 
\begin{eqnarray*}
 k^{-}_{1,2} &=& -i\gamma \pm\sqrt{\Omega^2-\gamma^2} \equiv k_{1,2}\,,
 \\
 k^{+}_{1,2} &=& +i\gamma \mp\sqrt{\Omega^2-\gamma^2} \equiv -k^{-}_{1,2} \,,
\end{eqnarray*}
 and 
$Amplitude_q(k)^2$ 
 is decreasing as 
$ |k| \to \infty $ . 
 Thus we expect, the maximal value of 
$amplitude_q(k)^2$
 is somewhere near the poles of 
$Amplitude_q(k)^2$,  
 at least at small 
$\gamma$ . 

 \addvspace{\bigskipamount}
 ...We have yet one analytic extension defined on 
${\bf C}$ except for some poles.
 It is 
$Det_{-1}(k)$, introduced in the beginning of this section:
\begin{eqnarray*}
 Det_{-1}(k) 
&:=&
 -\frac{1}
{ k^2+{2i\gamma} k - \Omega^2 }
 \,,\,\qquad  
 k^2+{2i\gamma} k - \Omega^2 \not=0 \,.
\end{eqnarray*}
 The expressions of 
$Amplitude$-s
 and 
$Det_{-1}(k)$
 suggest writing
$$
  Amplitude_q(k)^2 = \Omega^4 A_{mp}^2\,Det_{-1}(k)\,Det_{-1}(-k) \,,
$$
$$
  Amplitude_Q(k)^2 = (-k^2+\Omega^2)^2 A_{mp}^2\,Det_{-1}(k)\,Det_{-1}(-k) \,,
$$
$$
  Amplitude_{qQ}(k)^2 = (k^2)^2 A_{mp}^2\,Det_{-1}(k)\,Det_{-1}(-k) \,.
$$
 These relations are mathamatically correct. 
 Therefore, let us discuss 
$amplitude$-s connecting them with 
$Det_{-1}(k)$.

 We notice that 
$Det_{-1}(k)$ 
 is analytic extension of 
$det_{-1,+}(k)$ , 
 thus 
$Det_{-1}(k)$ 
 as well as 
$det_{-1,+}(k)$ , 
 is related rather to the resolvent,
$R(z)$, 
 resolvent as a function
\footnote{ operator-valued function }
 of parameter 
$z, z\in {\bf C}$
\footnote{
 we mean usual notations in usual writing 
$R(z)=(z-A)^{-1}$
 }. 
 But the `natural' parameter of 
 resolvent is 
$z$ ,
 not 
$k$ .
 Thus we would like to reformulate 
$Det_{-1}(k)$ 
 as a function of 
$z$ .

 \addvspace{\medskipamount}

 Let us resume what we have. 
 The situation with resolvent and resonances 
 discussed in this section looks generally like this:

\begin{Observation}{ 1. }

 (1) 
 The resolvent is related to the function 
\begin{eqnarray*}
 det_{-1,+}(k)
 &:=&
 -\frac{1}
{ k^2+2i\gamma k - \Omega^2}
 = 
  \frac{1}
{ (k+i\gamma)^2 - \Omega^2 +\gamma^2} \,,
\\&&\qquad\qquad\qquad
 \quad (\mbox{ where } Im\,k > 0) \,;\\
\end{eqnarray*}

 (1a) 
 the domain of 
$det_{-1,+}(k)$
 is 
$$
 \{k| Im\,k > 0\}
$$ 
 and is one-to-one to the set
$$
 \{z|z=-k^2; Im\,k > 0\} \equiv {\bf C}\setminus (-\infty,0] \,;
$$ 

 \addvspace{\medskipamount}

 (2) 
 The 
$amplitude$-s 
  are related to the function 
\begin{eqnarray*}
 det_{-1,0}(k)
 &:=&
 -\frac{1}
{ k^2+2i\gamma k - \Omega^2}
 = 
  \frac{1}
{ (k+i\gamma)^2 - \Omega^2 +\gamma^2} \,,
\\&&\qquad\qquad\qquad
 \quad (\mbox{ where } Im\,k = 0) \,;\\
\end{eqnarray*}

 (2a) 
 the domain of 
$det_{-1,0}(k)$
 is 
$$
 \{k| Im\,k = 0\} 
 \equiv 
 \{k| Im\,k = 0\,,\,Re\,k > 0 \} 
 \cup 
 \{k| \,k = 0\, \} 
 \cup
 \{k| Im\,k = 0\,,\,Re\,k < 0 \} \,,
$$ 
 with the possible exception of the poles
\footnote{ they exist, iff 
$\gamma = 0$} . 

 Note, 
$$
 \{z|z=-k^2; Im\,k = 0\,,\,Re\,k > 0 \} 
 \equiv (-\infty,0) 
 \equiv 
 \{z|z=-k^2; Im\,k = 0\,,\,Re\,k < 0 \} 
$$ 

 \addvspace{\medskipamount}

 (3) 
 The resonances are related to the poles of the function 
\begin{eqnarray*}
 det_{-1,-,0}(k)
 &:=&
 -\frac{1}
{ k^2+2i\gamma k - \Omega^2}
 = 
  \frac{1}
{ (k+i\gamma)^2 - \Omega^2 +\gamma^2} \,,
\\&&\qquad\qquad\qquad
 \quad (\mbox{ where } 
 Im\,k \leq 0\,,\,
 k^2+2i\gamma k - \Omega^2 \not= 0 ) \,;\\
\end{eqnarray*}

 (3a) 
  the poles, 
$k_{pol}$, 
 of 
$det_{-1,-,0}(k)$,  
 and resonant values 
 are connected by:
$$ 
 k_{pol} = -i\gamma \pm i\sqrt{- \Omega^2 +\gamma^2} \,,
$$
$$
 k_{1st.res} = \pm\sqrt{\Omega^2 - 2\gamma^2}\,, 
$$
$$
 k_{2nd.res} = \pm\Omega \,,
$$
$$
 k_{3rd.res} = \pm\sqrt{\frac{\Omega^4}{\Omega^2 - 2\gamma^2}} \,.
$$

 \addvspace{\bigskipamount}

 (3b) 
 The domain of 
$det_{-1,-,0}(k)$
 is 
$$
  \{k| Im\,k \leq 0\} 
   = \{k| Im\,k < 0\}\cup \{k| Im\,k = 0\} \,,
$$
 with the exception of the poles;

 the set 
$$
 \{k| Im\,k < 0\}
$$ 
 is one-to-one to the set
$$
 \{z|z=-k^2; Im\,k < 0\} \equiv {\bf C}\setminus (-\infty,0] \,;
$$

 \addvspace{\medskipamount}

 (4) 
 All these functions are consistent: 
 there is an analytic function,---
\begin{eqnarray*}
 Det_{-1}(k) 
 &=&
 -\frac{1}
{ k^2+{2i\gamma} k - \Omega^2 }
 \,,\,\qquad  
 k^2+{2i\gamma} k - \Omega^2 \not=0 \,,
\end{eqnarray*}
 --- 
 such that:

 (4a) 
 every 
$det_{-1,\#}$ 
 is a restricton of 
$ Det_{-1}(k) $ ;

 (4b) 
 the union of the domains of 
$det_{-1,\#}$ 
 is the domain of 
$ Det_{-1}(k) $ .

\end{Observation}

 \addvspace{\bigskipamount}

  Now then, we have no problem with reformulating 
$det_{-1,+}(k)$ 
 as a function of 
$z$: 
 we do it by defining 
\begin{eqnarray*}
 d_{-1,+}(z)
 &:=& 
 det_{-1}(k(z))
 =
 -\frac{1}
{ k(z)^2+2i\gamma k(z) - \Omega^2}
 = 
  \frac{1}
{ (k(z)+i\gamma)^2 - \Omega^2 +\gamma^2} \,,
\\&&{}
  \mbox{ if  } z\not\in (-\infty,0] \,,
\end{eqnarray*}
 where 
$k(z)$ is defined as the unique solution to  
$$
 k^2=-z \,,\, Im\,k > 0 \,.\\
$$ 

 We need now reformulate 
$Det_{-1}(k)$
 which is an extension of 
$det_{-1}(k)$. 
 In other words, we have to extend 
$d_{-1,+}(z)$ , 
 but `into WHAT?' 

 Of course, we can extend 
$d_{-1}(z)$ 
 onto
$$
 z \in {\bf C} \,,\, z=-k^2\,,\, k^2+2i\gamma k - \Omega^2 \not= 0 \,. 
$$
 We can do it, by defining, e.g., 
\begin{eqnarray*}
 d_{-1(1)}(z)
 &:=& 
 -\frac{1}
{ k_{+}(z)^2+2i\gamma k_{+}(z) - \Omega^2}
 =
  \frac{1}
{ (k_{+}(z)+i\gamma)^2 - \Omega^2 +\gamma^2} \,,
 \\&&{} 
( k_{+}(z)^2+2i\gamma k_{+}(z) - \Omega^2 \not=0 ) 
\end{eqnarray*}
 where 
$k_{+}(z)$ 
 is defined as the unique solution to  
$$
 k_{+}^2=-z \,,\, Im\,k_{+} > 0 \,,\\
$$ 
 if 
$$
 z\not\in (-\infty,0] \,, 
$$
 and to 
$$
 k_{+}^2=-z \,,\, Re\,k_{+} \geq 0 \,,\\
$$ 
 if 
$$
 z\in (-\infty,0] \,, 
$$
 but further, there is `no place' for the `rest of 
$Det_{-1}$', 
 i.e., for  
$det_{-1,-,0}$ .

 With 
$det_{-1,-,0}$ , 
 we connect 
 another function of 
$z$ : 
\begin{eqnarray*}
 d_{-1(2)}(z)
 &:=& 
 -\frac{1}
{ k_{-}(z)^2+2i\gamma k_{-}(z) - \Omega^2}
 =
  \frac{1}
{ (k_{-}(z)+i\gamma)^2 - \Omega^2 +\gamma^2} \,,
 \\&&{} 
( k_{-}(z)^2+2i\gamma k_{-}(z) - \Omega^2 \not=0 ) 
\end{eqnarray*}
 where 
$k_{-}(z)$ is defined as the unique solution to  
$$
 k_{-}^2=-z \,,\, Im\,k_{-} < 0 \,,\\
$$ 
 if 
$$
 z\not\in (-\infty,0] \,, 
$$
 and to 
$$
 k_{-}^2=-z \,,\, Re\,k_{-} \leq 0 \,,\\
$$ 
 if 
$$
 z\in (-\infty,0] \,. 
$$

 \addvspace{\bigskipamount}

 Thus we are not able to reformulate 
$Det_{-1}(k)$
 in terms of ONE function of
$z$ . 
 We can only reformulate  
$Det_{-1}(k)$
 in terms of TWO functions of ONE variable: 
$ d_{-1(1)}(z) $ 
 and 
$ d_{-1(2)}(z) $ .

 \addvspace{\bigskipamount}
 Nevertheless, 
 physicists prefer to say, 
 there is ONE analytic function 
$D_{-1}(z)$
 defined on `TWO exemplars' of 
${\bf C}$ , 
 so that

 \addvspace{\bigskipamount}

 (1) on the first `exemplar' of
$\bf C$,
$D_{-1}(z) =d_{-1(1)}(z)$ ; 
 this first `exemplar' of
$\bf C$
 is called {\bf first} or {\bf physical sheet} of  
$F(z)$ ;

 (2) on the second `exemplar' of
$\bf C$,
$D_{-1}(z) =d_{-1(2)}(z)$ ; 
 this second `exemplar' of
$\bf C$
 is called {\bf second} or {\bf unphysical sheet} of  
$F(z)$ .

 \addvspace{\bigskipamount}

 One can say also: 

 \addvspace{\bigskipamount}

$d_{-1(2)}(z)$ 
 is a part of an {\bf extended} 
$d_{-1(1)}(z)$, the part which is defined on the second `exemplar' of 
$\bf C$,   
 and this second `exemplar' of 
$\bf C$  
 is named the {\bf second sheet}  
 associated with 
$R(z)$. 

 \addvspace{\bigskipamount}

 ---In these words,  
$z_{res}$ 
 is a {\bf resonant pole},
 if 
$z_{res}$ 
 is a pole placed on the second sheet of the resolvent. 

 \addvspace{\medskipamount}

 From the formal mathematical standpoint, 
 the notion we have just introduced can be explained as following: 
`two exemplars' of 
${\bf C}$
 can be defined as 
$$
 ({\bf C}\times \{1\})\cup({\bf C}\times \{2\}) \,; 
$$
 the `first' sheet is identified with 
${\bf C}\times \{1\}$ , 
 and the second one is identified with  
${\bf C}\times \{2\}$; 
 finally put 
$$D_{-1}(z\times \{1\}):= d_{-1(1)}(z) \,, $$
$$D_{-1}(z\times \{2\}):= d_{-1(2)}(z) \,. $$

 \addvspace{\medskipamount}

 The constructions of such kind are basic objects of 
 the theory of {\bf Riemann surfaces}, but we will not go into details.  

 There are reasons for it, and one of them is that 
 the concept of Riemann surface has been elaborated 
 so as to 
 avoid introducing principal distinctions between the sheets, 
 whereas we would like to contradistinguish them,---we need two or even three 
 objects,---one object connected immediately with resolvent 
 and the other(s) one(s) done 
 with amplitudes and resonant poles. 
 In other words, we prefer constructions like that we have displayed 
 as {\bf Observation 1 }.

 Another reason is that we need rather 
$R(z)$ 
 and 
$R(-k^2)$ than 
$det_{\#}$ , 
 but the formers 
 are not usual scalar functions. They are operator-valued function,
 thus, we must give exact definitions suited to the case. 
 First we must define what we mean by `element of
$R(-k^2)$ 
 is an analytic function', 
 what we mean by `analytic extension', 
 and then many other things. 
 Otherwise confusion and false conclusions will occur!!

 \addvspace{\bigskipamount}

 How to find proper mathematical definitions, 
 does not form the subject of this paper.

 \addvspace{\bigskipamount}

\newpage\section*%
{ Appendix A. }  
\section*%
{  Free Green's Function. }

 Here we reproduce a well-known proof that

\begin{eqnarray*}
 (z-c^2\Delta)^{-1}(x,x') 
&=&
 \frac{i}{2k c^2}e^{ik|x-x'|}  ,
 \mbox{ where }  \quad k^2=-z/c^2, \quad Im\,k > 0 \\
\end{eqnarray*}

 It can be deduced from p-representation of 
$(z - c^2\Delta)^{-1}$:
$$
 (z-c^2\Delta)^{-1}(x,x') =
 \frac{1}{2\pi}\int_{-\infty}^{+\infty}
 \frac{e^{ip(x-x')}}{z+c^2p^2} dp \,.
$$

 We have: 
\begin{eqnarray*}
 (z-c^2\Delta)^{-1}(x,x') 
 &=&
 \frac{1}{2\pi}\int_{-\infty}^{\infty}\frac{e^{ip(x-x')}}{z+c^2p^2} dp
\\
 &=&
 \frac{1}{c^2}
 \frac{1}{2\pi}\int_{-\infty}^{\infty}\frac{e^{ip(x-x')}}{p^2+z/c^2} dp
\\
 &=&
 \frac{i}{c^2}
 \frac{1}{2\pi i}\int_{-\infty}^{\infty}\frac{e^{ip(x-x')}}{p^2-k^2} dp
\\
 &=&
 \frac{i}{c^2}
 \frac{1}{2\pi i}\int_{-\infty}^{\infty}
   \frac{e^{ip(x-x')}}{p+k}\frac{1}{p-k} dp
\\[\smallskipamount]\\
 &=& 
 \frac{i}{c^2}
 \frac{e^{ik|x-x'|}}{2k}  ,
\\[\smallskipamount]\\
 &=& 
 \frac{i}{2k c^2}e^{ik|x-x'|}  ,
 \mbox{ where }  \quad k^2=-z/c^2, \quad Im\,k > 0 \\
\end{eqnarray*}

\newpage 

\newpage 

\bibliographystyle{unsrt}

\begin{thebibliography}{DalKre}
\bibitem[AK]{AK} 
 {\sc S. Albeverio and P. Kurasov}, 
 {\it Singular Perturbations of Differential Operators. 
 Solvable Scr\"odinger Type Operators}, 
 London Mathematical Society: 
 Lecture Note Series. 271, 1999.
 {\sc Cambridge university press}   


\bibitem[BF]{BF} 
 {\sc F.A. Berezin, L.D. Faddeev}, 
 {\it Remark on the Schr\"odinger equation with singular potential}, 
 Dokl. Akad. Nauk. SSSR, 137 (1961) 1011-1014 (in Russian).

\bibitem[Do]{Do} 
 {\sc W. Donoghue}, 
 {\it On the perturbation of spectra},
 Comm. Pure App. Math. 18 (1965) 559-579


\bibitem[Fog]{Fog} 
 {\sc S.R. Foguel}, 
 {\it Finite Dimensional Perturbations In Banach Spaces}, 
 American Jornal of Mathematics, 
 Volume 82, Issue 2 ( Apr., 1960 ), 260-270

\bibitem[Fr]{Fr} 
 {\sc K.O. Friedrichs},
 {\it Perturbation of Spectra in Hilbert Space},
 American Mathematical Society, Providence, (1965)


\bibitem[Gan]{Gan} 
 {\sc F.R. Gantmaher},
 {\it The theory of matrices.} 4-th ed.., Moscow,
 Nauka (1988)(Russian);
 English  translation: Chelsea publishing company,
 New York (1959).


\bibitem[Jack]{Jack}
 {\sc J.D.~Jackson:} {\it Classical electrodynamics}.
\\  John Wiley \& Sons, Inc. New York-London, 1962.
\smallskip\\
 see `Radiation Reaction, Abraham Lorentz Equation, Braking Radiation'
 and all that. 


\bibitem[RS1]{RS1}
 {\sc M. Reed, B. Simon}, 
 {\it Methods of Modern Mathematical Physics, vol 1, 
  Functional analysis}, 
   - N.Y.: Academic Press, 1972.


\bibitem[RS2]{RS2}
 {\sc M.~Reed, B.~Simon}, 
 {\it Methods of Modern Mathematical Physics, vol 2, 
 Fourier analysis, Self-Adjointness}, 
 - N.Y.: Academic Press, 1975.

\bibitem[RS3]{RS3}
 {\sc M. Reed, B. Simon}, 
 {\it Methods of Modern Mathematical Physics, vol 3, 
 Scattering Theory}, - N.Y.: Academic Press, 1979.

\bibitem[RS4]{RS4}
 {\sc M. Reed, B. Simon}, 
 {\it Methods of Modern Mathematical Physics, vol 4,
 Analysis of Operators}, 
 - N.Y.: Academic Press, 1978.

\bigskip 

 {\large \bf Electronic Print: }

\bigskip

 {\large \bf Mathematical Physics Preprint Archive  }

\bibitem[DerFr]{DerFr}
 mp\_arc 02-275

 Derezinski J., Fruboes R.

 Renormalization of the Friedrichs Hamiltonian

(16K, LATeX 2e)
\begin{verbatim}
 http://www.ma.utexas.edu/mp_arc-bin/mpa?yn=02-275
 http://mpej.unige.ch/mp_arc-bin/mpa?yn=02-275
 http://www.maia.ub.es/mp_arc-bin/mpa?yn=02-275
\end{verbatim}



\bibitem[Der]{Der}
 mp\_arc 02-300

 Jan Derezinski

 Van Hove Hamiltonians---exactly solvable models of the infrared
 and ultraviolet problem.

 (62K, LATeX 2e)
\begin{verbatim}
 http://www.ma.utexas.edu/mp_arc-bin/mpa?yn=02-300
 http://mpej.unige.ch/mp_arc-bin/mpa?yn=02-300
 http://www.maia.ub.es/mp_arc-bin/mpa?yn=02-300
\end{verbatim}

\bibitem[03-33]{03-33}
 mp\_arc 03-33

 Sergej A. Choroszavin ( sergius@pve.vsu.ru )

 An Interaction of An Oscillator with 
 An One-Dimensional Scalar Field. 
 Simple Exactly Solvable Models based on 
 Finite Rank Perturbations Methods. 
 I: D'Alembert-Kirchhoff-like formulae 

(88K, LaTeX 2.09) Jan 29, 03 
\begin{verbatim}
 http://www.ma.utexas.edu/mp_arc-bin/mpa?yn=03-33
 http://mpej.unige.ch/mp_arc-bin/mpa?yn=03-33
 http://www.maia.ub.es/mp_arc-bin/mpa?yn=03-33
 http://kleine.mat.uniroma3.it/mp_arc-bin/mpa?yn=03-33
\end{verbatim}




 {\large \bf arXiv.org e-Print archive (LANL E-Print) } 


\bibitem[AMN]{AMN}
 Paper: physics/0001009
\\ From: \verb"adolfo@lafexSu1.lafex.cbpf.br" (Adolfo Malbouisson)
\\ Date: Wed, 5 Jan 2000 19:20:08 GMT   (15kb)
\\[2ex]
 Title: 
 An Exact Approach to the Oscillator Radiation Process in an Arbitrarily
 Large Cavity
\\ Authors: N.P. Andion, A.P.C. Malbouisson and A. Mattos Neto
\\ Comments: 27 pages
\\ Subj-class: Atomic Physics; Mathematical Physics
\\
  \verb"http://arXiv.org/abs/physics/0001009" 

\bibitem[BKZ]{BKZ}
 Paper: math-ph/0210051
\\ From: Volker Bach \verb"<vbach@mathematik.uni-mainz.de>"
\\ Date: Wed, 30 Oct 2002 00:06:05 GMT   (31kb)
\\[2ex]
 Title: 
 Mathematical analysis of the photoelectric effect
\\ Authors: Volker Bach, Frederic Klopp and Heribert Zenk
\\ Comments: See also http://www.intlpress.com/ATMP
\\ Subj-class: Mathematical Physics
\\ Journal-ref: Adv. Theor. Math. Phys. 5 (2001) 969-999
\\ 
 \verb"http://arXiv.org/abs/math-ph/0210051" 

\bibitem[Ch03-1]{Ch03-1}
 Paper: math.DS/0301167
\\ From: "Sergej A. Choroszavin" \verb"<sergius@pve.vsu.ru>" 
\\ Date: Thu, 16 Jan 2003 04:34:16 GMT   (18kb)
\\ Date (revised v2): Fri, 17 Jan 2003 03:07:58 GMT   (18kb)
\\ Date (revised v3): Mon, 27 Jan 2003 19:00:54 GMT   (18kb)
\\[2ex]
 Title: 
  An Interaction of An Oscillator with An One-Dimensional Scalar Field.
  Simple Exactly Solvable Models based on Finite Rank Perturbations Methods. I:
  D'Alembert-Kirchhoff-like formulae
\\ Author: Sergej A. Choroszavin
\\ Comments: Latex 2.09 
\\Subj-class: Dynamical Systems; Mathematical Physics
\\
 \verb"http://arXiv.org/abs/math.DS/0301167" 

 This paper is the same as in   
 Mathematical Physics Preprint Archive, PAPER: 03-33 
\\


\bibitem[MPB]{MPB}
 Paper: hep-th/9207033
\\ From: physth@ulb.ac.be
\\ Date: Fri, 10 Jul 92 16:52:54 +0200   (20kb)
\\[2ex]
 Title: 
  On the Problem of the Uniformly Accelerated Oscillator
\\ Authors: S. Massar, R. Parentani, R. Brout
\\ Comments: 14 pages (+postscript figures attached), ULB-TH-03/92
\\
 \verb"http://arXiv.org/abs/hep-th/9207033" 




\end{thebibliography}

\end{document}